\documentclass[useAMS,usenatbib]{mn2e}
\usepackage{graphicx}
\usepackage{txfonts}
\usepackage{natbib}
\usepackage{xcolor}
\usepackage{ amssymb }
\def\gsim{\mathrel{\hbox{\rlap{\hbox{\lower4pt\hbox{$\sim$}}}\hbox{$>$}}}}
\def\lsim{\mathrel{\hbox{\rlap{\hbox{\lower4pt\hbox{$\sim$}}}\hbox{$<$}}}}

\begin{document}
   \title[Inclination-dependent spectral and timing properties]{Inclination-dependent spectral and timing properties in transient black hole X-ray binaries}

   \author[]{L.M.~Heil$^1$, P.~Uttley$^1$ and M. Klein-Wolt$^{2,3}$\\
$^{1}$Anton Pannekoek Institute, University of Amsterdam, Science Park 904, 1098 XH Amsterdam, The Netherlands \\
$^{2}$Department of Astrophysics, Research Institute of Mathematics, Astrophysics and Particle Physics, Radboud University Nijmegen,\\
Heijendaalseweg 135, 6525 AJ Nijmegen, The Netherlands\\
$^{3}$Science \& Technology, Olof Palmestraat 14, 2616 LR Delft, The Netherlands}

   \date{Draft \today}

   \pagerange{\pageref{firstpage}--\pageref{lastpage}} \pubyear{2002}

   \maketitle
   \label{firstpage}

\begin{abstract}
We use a simple one-dimensional parameterisation of timing properties to show that hard and hard-intermediate state transient black hole X-ray binaries with the same power-spectral shape have systematically harder X-ray power-law emission in higher-inclination systems.  We also show that the power-spectral shape and amplitude of the broadband noise (with low-frequency quasi-periodic oscillations, QPOs, removed) is independent of inclination, confirming that it is well-correlated with the intrinsic structure of the emitting regions and that the ``type C'' QPO, which is inclination-dependent, has a different origin to the noise, probably geometric.  Our findings suggest that the power-law emission originates in a corona which is flattened in the plane of the disc, and not in a jet-like structure which would lead to softer spectra at higher inclinations.  However, there is tentative evidence that the inclination-dependence of spectral shape breaks down deeper into the hard state. This suggests either a change in the coronal geometry and possible evidence for contribution from jet emission, or alternatively an even more optically thin flow in these states.
\end{abstract}

\section{Introduction}

Recently, \citet{Munoz-Darias13} showed that in the disc-dominated low mass black hole X-ray binary (BH LMXB) soft states, systems with higher inclination of the binary orbit to our line of sight appear systematically harder than lower-inclination objects. In the soft state there is a general consensus that the inner edge of the disc is close to the central black hole \citep{Gierlinski04}.  As a consequence, the observed disc blackbody spectra can be significantly altered by Doppler shifting and relativistic effects. The resulting spectral shape is expected to depend strongly on the viewing angle to the source, explaining the inclination-dependence of the soft state spectrum \citep{Munoz-Darias13}.

If inclination-dependent differences in emission can be seen in the hard and intermediate states, this would provide important clues about the geometry of the `coronal' X-ray power-law emitting regions (i.e. whether they are 'disc-like', or more like a jet).  An important distinction with the soft states is that, while the soft state spectra appear very stable, corresponding to a constant inner disc radius \citep{Gierlinski04,Steiner10}, the intermediate and hard states show strongly evolving spectral properties which may be linked to large changes in structure, e.g. due to truncation of the disc inner radius. Comparing different systems is therefore difficult, since it is hard to know which particular observations correspond to the same intrinsic emission geometry. We must make use of additional information in order to break the degeneracy between systematic inclination-dependent differences and real changes in emission geometry during an outburst.  The strong correlation of timing properties with spectral evolution \citep{Homan01,Belloni05} indicates that timing is a good proxy for the changes in structure of the emitting regions and hence may provide the extra information needed. We can compare objects with the same timing properties, assuming that they have the same emission geometry. To do so, we need a simple way of comparing the timing properties of different objects.  

In Heil, Uttley \& Klein-Wolt (2014), henceforth Paper~I, we demonstrate a new method for directly comparing power spectra from different objects. The power colour-colour diagram uses ratios of integrated power (i.e. variance) in different frequency bands to compare the shape of power spectra in a simple, model-independent way.  All BH LMXBs trace a consistent `loop' in the power colour-colour diagram as they go through an outburst.  We can parameterise the power-spectral shape from an observation by using the angular position of that observation around the loop in the power colour-colour diagram.  We call this angle the {\it hue}, by analogy with the `colour-wheel' concept of the same name.  As shown in Paper~I, similarly-shaped power spectra have the same hue.  In this letter we use the power spectral hue together with the spectral hardness to compare the energy spectral and timing properties of hard and intermediate state BH LMXBs in a model independent way, and so investigate their inclination-dependence.

\section{Data Analysis}

\begin{figure*}
\begin{center}

	\includegraphics[width=9.0 cm, angle=90]{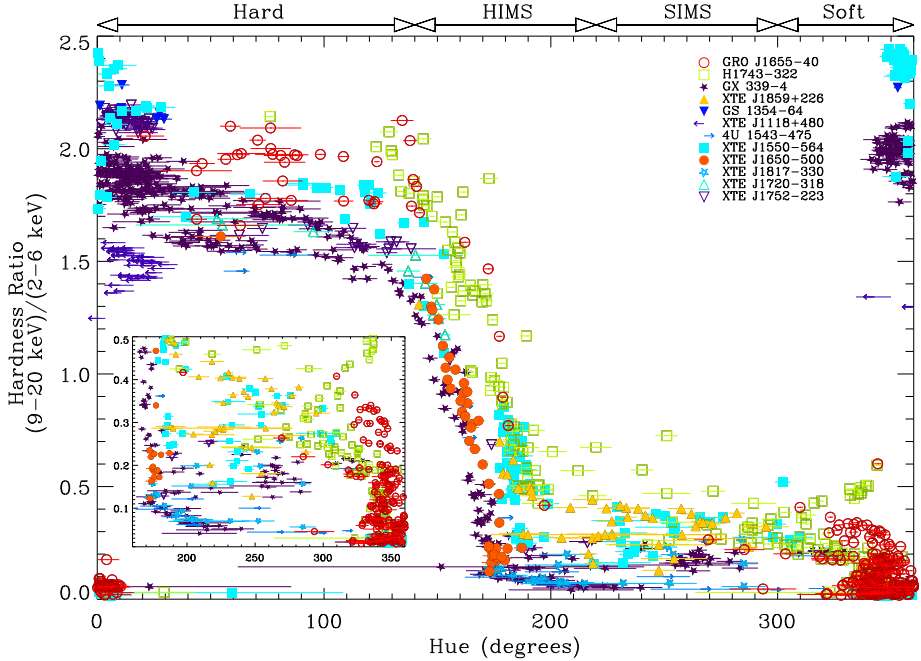}
\end{center}
\caption{Hardness vs. power-spectral hue for BH LMXBs. Broadly speaking, values of hue around 180$^{\circ}$ have power spectra which are the most `peaked', while hue values approaching 0$^{\circ}$ or 360$^{\circ}$ have much broader power spectra, corresponding to the hard and soft states respectively.  The ranges of hue corresponding to the different states are indicated at the top of the Figure.  Note that there is some overlap in hue between the soft and the hard state.}
\label{fig:hrdrmsvang}
\end{figure*}
The data analysis is discussed in greater detail in Paper~I. To briefly summarise, we take all observations of 12 BH LMXBs (given in Table \ref{tab:srcs}) taken using the Proportional Counter Array (PCA) on the {\it Rossi X-ray Timing Explorer} ({\it RXTE}). Power spectra were measured in the 2--13~keV energy range and then split into four frequency-bands each spanning a factor of 8 in frequency; they cover the following ranges 0.0039-0.031 Hz, 0.031-0.25 Hz, 0.25-2.0 Hz and 2.0-16.0 Hz. The Poisson noise was estimated through fitting and removed and the power spectra were integrated over each frequency band to give the variance contributed by each frequency range.  For clarity, observations where the variance was unconstrained at a 3$\sigma$ level in any frequency band were discarded from the sample.

Following Paper~I, we obtain the power colour-colour diagram by plotting on each axis the following ratios of variance: PC1=0.25-2.0 Hz/0.0039-0.031 Hz and PC2=0.031-0.25 Hz/2.0-16.0 Hz (see Fig.~2 of Paper~I and Fig.~\ref{fig:pwcolinc} in this paper for examples). Errors for the ratios are propagated in quadrature. We can assign an angle or `hue' to each point around the power-colour-colour diagram, by taking the dot product between the semi-major axis of the power colour-colour loop and the vector from the centre of mass to the measured power-colour value (see Paper~I, Section~3.1 for further details).  The hue therefore provides a way to classify power-spectral shape using a single variable.

Evolution around the power-colour-colour diagram is initially in a clockwise direction from the line in Figure \ref{fig:pwcolinc}a where power spectral shapes are broad with roughly equal amounts of power in each frequency band (i.e. both power-colour ratios are $\sim$1.0). Starting from hue 0$^{\circ}$ objects transition through the hard state clockwise along the upper path before reaching the hard-intermediate state, found in the lower-right hand corner of the power-colour-colour diagram (hue $\sim$180$^{\circ}$). As sources evolve though the SIMS and into the soft states they continue to track along a clock-wise path, returning to a hue $\sim$0$^{\circ}$. Evolution back into the hard state follows the same loop only in an anti-clockwise direction.

In order to measure the energy spectral-hardness independently of long-term changes in the PCA instrument response, fluxes are generated in a model-independent way by dividing the PCA Standard 2 mode spectrum by the effective area of the instrument response in each spectral channel.  This is carried out by unfolding the spectrum with respect to a zero-slope power-law (i.e. a constant) in the {\sc xspec} spectral-fitting software, and measuring the unfolded flux over the specified energy range (interpolating where the specified energy does not fall neatly at the edge of a spectral channel).

\section{Results}
\subsection{Inclination-dependence of spectral hardness}

\begin{table}
\centering
\begin{tabular}{lrrrr}
\hline\hline
Source Name &  Total Obs. & Good Obs. & Inc. Ang. & Group \\
(1)      & (2) & (3) & (4) & (5) \\

\hline
GX 339-4 & 619 & 457 & $\lesssim$ 60$^\circ$ $^1$ & low \\
XTE J1118+480 & 65 & 48 & 68 $\pm$ 2$^\circ$ $^2$ & high\\
GS 1354-64 & 7 & 7 & - & - \\ 
4U 1543-475 & 51 & 20 & 20.7 $\pm$ 1.5$^\circ$ $^3$ & low \\
XTE J1550-564 & 268 & 201 & 74.7 $\pm$ 3.8$^\circ$ $^4$ & high \\
XTE J1650-500 & 97 & 41 & $\gtrsim$47$^\circ$ $^5$ & low \\
GRO J1655-40 & 438 & 303 & 70.2 $\pm$ 1$^\circ$ $^6$ & high \\
XTE J1720-318 & 85 & 10 & - & - \\
H 1743-322 & 451 & 177 & 75 $\pm$ 3$^\circ$ $^8$ & high \\
XTE J1752-223 & 143 & 64 & $\lesssim$ 49$^\circ$ $^{9}$ & low \\
XTE J1817-330 & 129 & 26 & - & - \\
XTE J1859+226 & 107 & 44 & $\gtrsim$ 60$^\circ$ $^{10}$ & - \\
\hline
Total & 2460 & 1398 \\
\hline
\multicolumn{3}{c}{{\it }}\\
\end{tabular}
\caption{BH LMXB sample:  (1) Source Name; (2) Total number of observations in the sample; (3) ``Good'' observations, i.e. those where the variance is constrained as non-zero at a 3 $\sigma$ level in all frequency bands. Observations in the low flux hard state or the soft state may be combined and are then counted as a single observation, these were grouped when they occurred within 5 days of one another, the hardness differed by no more than 5\% and there were no obvious differences in the power spectra based on a visual inspection; (4) Measured inclination angles for the systems; (5) Assigned inclination angle group. References: $^1${\protect\cite{Zdziarski98}}; $^2${\protect\cite{Gelino06}}; $^3${\protect\cite{Orosz03}}; $^4${\protect\cite{Orosz11a}};  $^5${\protect\cite{Orosz04}}; $^6${\protect\cite{Greene01}}; $^7${\protect\cite{Kuulkers13}}; $^8${\protect\cite{Steiner12}}; $^9${\protect\cite{Miller-Jones11}}; $^{10}${\protect\cite{Corral-Santana11}}; $^{11}${\protect\cite{Orosz11b}}}
\label{tab:srcs}
\end{table}

\begin{figure}
\begin{center}
	\includegraphics[width=8.0 cm, angle=0]{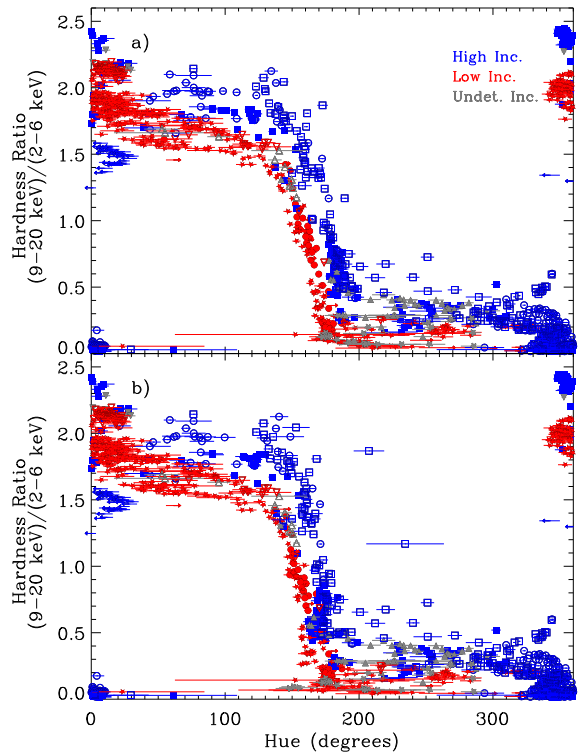}
\end{center}
\vspace{-0.3cm}
\caption{Hardness vs. power-spectral hue, colour-coded according to inclination angle. The lower panel is identical but uses power spectra with the QPOs removed to calculate their hue.  Note that the outlying hard values around hue$\sim 200^{\circ}$ in the lower panel are due to unusual low-frequency QPOs in 1H~1743-322, which are not removed from the power spectra (see Paper~I and \citealt{Altamirano12}). Objects with no measured inclination angle are coloured black.}
\label{fig:hrdvang}
\end{figure}

\begin{figure}
\begin{center}

	\includegraphics[width=8.0 cm, angle=0]{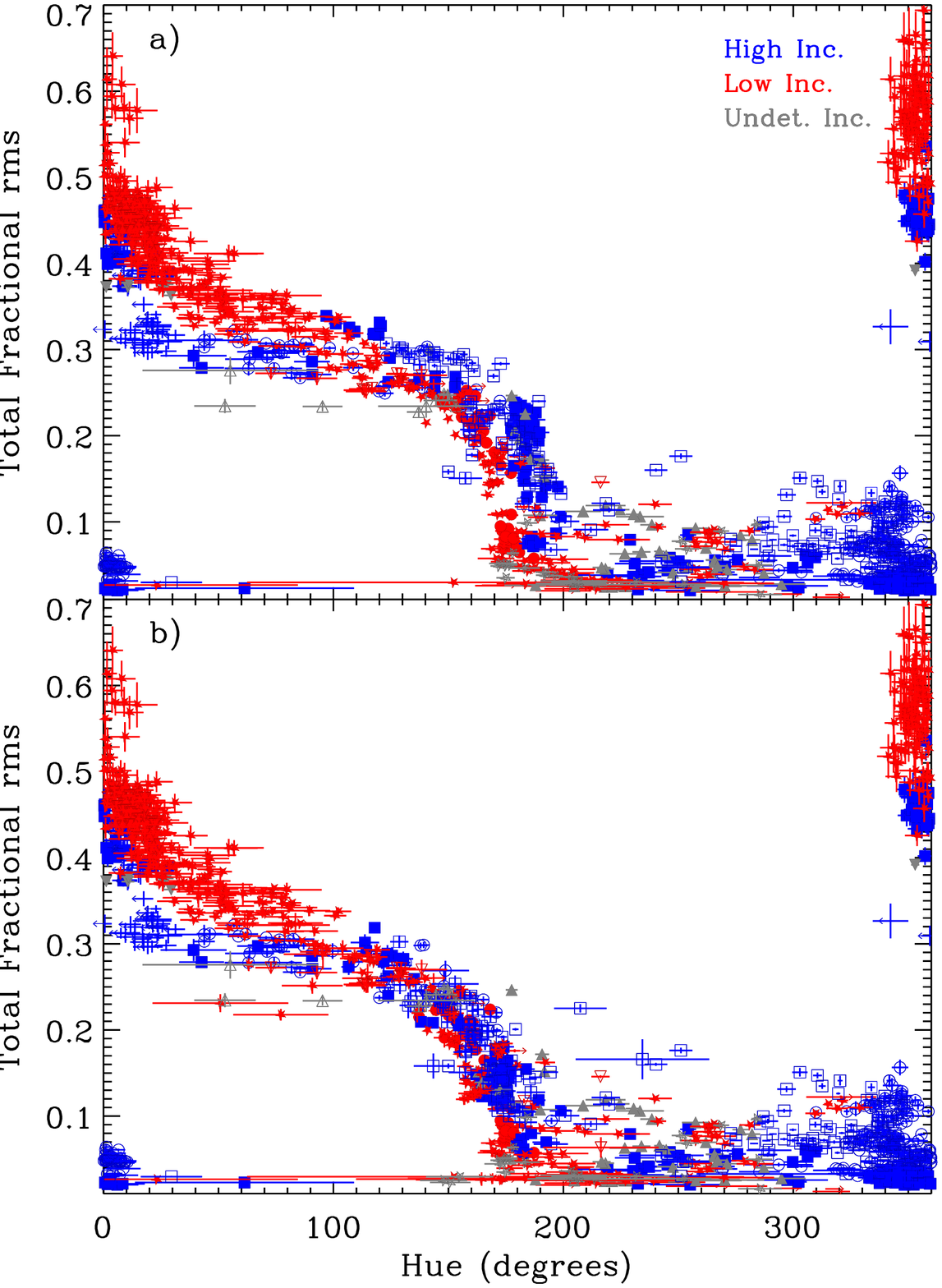}
\end{center}
\vspace{-0.3cm}
\caption{Total 0.0039--16~Hz rms vs. hue colour coded according to inclination angle, using the full power-spectrum (a) and broadband noise only, with QPOs removed (b).}
\label{fig:rmsincqporm}
\end{figure}
In Figure \ref{fig:hrdrmsvang} we plot the hue against energy-spectral hardness. We observe clear evolution of the hardness with the hue, as expected from the known correlation of timing properties and energy-spectral shape \citep{Homan01,Belloni05}.  However, there is also considerable spread between the paths followed by different objects observed in the hardness for similar hues, i.e. similar power-spectral shapes. Is present in part of the hard states and all of the hard intermediate (HIMS) and soft intermediate (SIMS) states, covering the hue ranges from 40--300 degrees (see Paper~I). Still, given that individual objects do appear to follow quite well-constrained paths - this suggests that it is the properties of each system which determines the observed spread. 

Following \cite{Munoz-Darias13} we use the inclination angle of the objects in our sample to classify the sources into two groups (high or low binary inclination), in order to explore whether this causes the observed spread in hardness in Figure \ref{fig:hrdrmsvang}. For consistency we classify into high or low inclination angle groups in the following manner: for the high inclination group we require both strong constraints on the measured inclination angle (when it is $>$60$^{\circ}$)  and physical evidence of high inclination in the emission in the form of absorption dips or equatorial winds \citep{McClintock03, Homan05, Kuulkers98, Ponti12}. This means that the objects follow the same classification scheme as \cite{Munoz-Darias13}, but XTE~J1859+226 is unclassified as it has an inclination angle constrained to be $>$60$^{\circ}$ but no other physical indicators have been observed. We note that although the inclination angle for XTE J1650-500 is measured to be $>$47$^{\circ}$, \cite{Ponti12} found only well constrained upper limits on the Fe XXVI absorption line in the soft state. This suggests that there is little evidence for equatorial disc winds observed from high inclination sources in this state, thus we classify it in the low inclination angle group. Table~\ref{tab:srcs} shows the most recently measured inclination angles and our classification for each object.

 Figure \ref{fig:hrdvang}a shows the hardness vs. hue plot colour coded according to inclination angle, it demonstrates that there is a clear inclination-dependence of the hardness of the track followed (note that discussion of the lower panel can be found later in Section \ref{sec:qpo}). The result is consistent with that observed by \cite{Munoz-Darias13} for the soft states: higher inclination objects with similar power-spectral hue (and thus similar power-spectral shape) are consistently harder than those at lower inclinations. Furthermore, we see that this inclination dependence extends through the SIMS, HIMS and into the hard state, where no significant disc emission is present in the {\it RXTE} PCA bandpass. For the lowest hue values (i.e. deeper into the hard state), the inclination-dependence of hardness appears to break down, with a much wider range of hardness values for both low and high-inclination sources at the same hue. The deep hard state is characterised by particularly broad power spectra (i.e. where both power-colour ratios are close to 1.0) and source luminosities $\lesssim$10$^{36}$ erg s$^{-1}$. It corresponds to low flux observations where the energy spectra begin to soften slightly \citep[see][for further discussion]{Wu08, Coriat11}, and this additional spectral change causes the apparent merging in Figure \ref{fig:hrdvang}.

\subsection{Fractional rms and the effect of quasi-periodic oscillations}
\label{sec:qpo}
We also search for any correlation between total fractional rms (in the 0.039-16.0 Hz frequency band) and inclination angle. Figure \ref{fig:rmsincqporm} shows that objects from the two inclination angle groups follow particular paths, although there is more mixing than that observed in the hardness-hue plot. There is notably a significant spread in hue for a particular rms value. This may be of some concern as it suggests that similar hue may not directly represent similar power-spectral shape. 

\begin{figure}
\begin{center}
	\includegraphics[width=8.0 cm, angle=0]{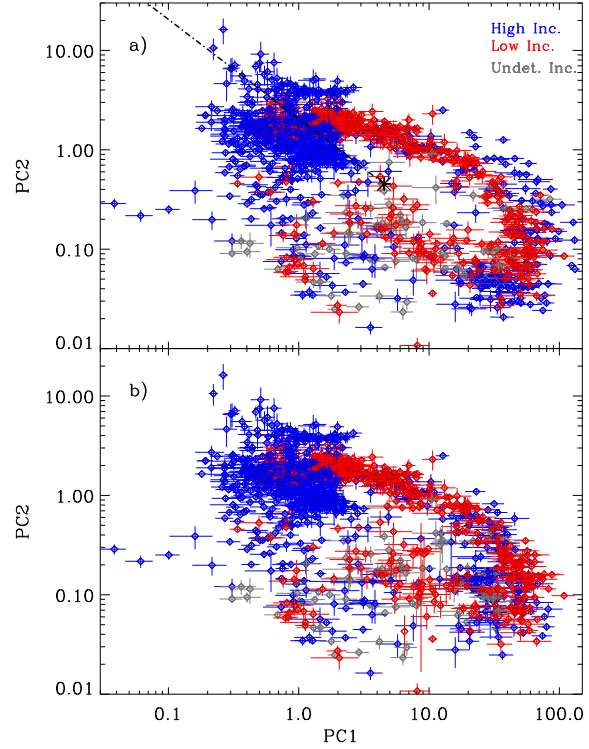}
\end{center}
\vspace{-0.5cm}
\caption{Power-colour-colour plots colour coded according to inclination angle. Differences in position for objects with different inclination angles are clearly visible. Panel b) shows the same figure made from power spectra with the QPOs removed, many of the major differences are no longer visible.}
\label{fig:pwcolinc}
\end{figure}

Previous work has revealed that most of these objects show strong quasi-periodic oscillations (QPOs) within their lightcurves, particularly in the hard state, HIMS and SIMS \citep[e.g. see review by][]{Belloni10}. These features are typically stronger in higher-inclination objects than lower and, within an outburst from a single object, are stronger on the rise into the outburst than in the return \citep{Schnittman06,Motta14}. As these QPOs contribute a significant amount of power to the lightcurve (up to $\sim$20$\%$), it is natural to assume that they must have some effect on the observed power-colour-colour diagram and the hue. In order to test for these differences within the power-colour-colour diagram itself we simply colour code all points on the diagram for inclination angle, using the same definitions as those in Figure \ref{fig:hrdvang}. In Figure \ref{fig:pwcolinc} we can clearly see the effect of the stronger quasi-periodic features in the HIMS of the high-inclination sources. The high-inclination sources in the hard state extend further to the right in the PC1 direction, while in the HIMS they extend lower in the PC2 direction than the low-inclination sources.  This effect is due to the combination of greater QPO amplitude at high-inclinations and the increase in QPO frequency from the 0.25--2~Hz to the  2--16 Hz band.

We therefore attempt to correct the power-colours and hue measurements for the presence of the QPOs. To do this, the contribution of the features (fundamental QPO plus any additional harmonics or subharmonics) is evaluated through fitting and then directly removed from the power spectra. Note that we only remove the type C QPOs \citep{Casella05}, i.e. from the hard state and HIMS, and do not consider the type B QPOs in the SIMS, which are generally weaker and less well-defined (but see \citealt{Motta14} for a discussion of the type B QPO inclination-dependence).  

Following \cite{Nowak00} we fit the power-spectra with a series of Lorentzians representing both broad and narrow features. An initial guess at the number of Lorentzians required is found through evaluating the hue. The broad-band noise is fitted with three broad Lorentzians \citep{Nowak00, KleinWolt08}, where QPOs are expected to be present three additional Lorentzian components are added. $\sim$65 intermediate soft states from H1743-322 and GRO J1655-40 are also found to require two additional Lorentzian components over the three used for typical soft state observations \citep{Motta12}. All additional components over those of the initial broad-band noise are tested for significance using an F-test, if the F-test probability is less than 0.001, then the QPO features are assumed to be required to fit the power-spectra. For power-spectra where QPOs are evaluated as significant features we remove the power in Lorentzians fitted to the QPOs from the power-spectra and then measure the power in each broad frequency band again.  Errors on the model QPO variance are accounted for by scaling the model variance by the error on the fitted Lorentzian normalisation, these are then propagated through with the measured error on the variance in the relevant frequency band. The power colour-colour diagram with the QPOs removed is plotted in Figure \ref{fig:pwcolinc}b. We find there is a maximum shift in hue of $\sim$ 20$^{\circ}$ for observations with strong type-C QPOs, removing the majority of the differences in hue between the high and low inclination objects in the hard state and the HIMS. This demonstrates that high and low inclination objects track a slightly different path around the power-colour diagram due to the differences in strengths of the QPOs. It is important to note however, that the weaker quasi-periodic features observed on the return to quiescence for all objects mean that there is significant overlap between objects at all inclinations. 


In order to test whether the QPO inclination dependence affects our conclusion that spectral hardness is inclination-dependent, we re-plot the hardness-hue and rms-hue plots but this time using values for the hue and rms which are measured from the power-colour-colour diagram with the QPOs removed. As Figure \ref{fig:rmsincqporm}b shows, the rms-hue plot is now much more consistent, to the point where there now appears to be no, or very little, clear difference between high and low inclination objects.  The hardness-hue plot (Figure \ref{fig:hrdvang}b) still shows clear differences in the hardness between objects at different inclinations. 

\section{Discussion}

To summarise, our key findings are:
\begin{enumerate}
\item When comparing sources with the same power-spectral shape, as defined by the hue, BH LMXBs in all states except the deepest (lowest hue) hard state show systematically harder X-ray spectra at higher system inclinations.
\item The inclination-dependence of QPO amplitude causes an offset between high and low-inclination sources in the HIMS part of the power-colour-colour diagram. Removing the QPO contribution does not change our conclusion about the inclination-dependence of spectral hardness.
\item The fractional rms of the broadband noise (i.e. with any QPOs removed) shows no evidence for inclination-dependence.
\end{enumerate}
\cite{Munoz-Darias13} find clear differences in the hardness values within the soft state and they interpret these in terms of relativistic effects on the disc spectrum. However, we find that this relation also exists in the HIMS and the hard state, which suggests that there is also a significant difference in the power-law energy spectral component.  We note that \cite{Munoz-Darias13}, find no clear energy spectral power-law dependence on inclination angle. This is likely to be due to the difficulty of controlling for the degeneracy between inclination effects and the changes in emission geometry from the hard to the soft state.  Selecting on the timing properties using the hue breaks this degeneracy, implying that the broadband power-spectral shape is a good tracer of the geometry of the emitting regions.

We can also consider that the split between different inclination angle groups may be due to inclination-dependent power-spectral differences rather than inclination-dependent energy spectral differences. However, we think this is unlikely as on the diagram both high and low-inclination sources show the initial hard-soft spectral transition (the downturn in hardness) at similar values of hue, and the same is true for the flattening of the hardness-hue trend on entry to the soft state. The simplest explanation for our result is that higher-inclination objects are systematically harder in the HIMS (continuing into the SIMS) and at least part of the hard state.  

To verify that the inclination-dependence of spectral hardness in the hard state and HIMS is associated with the power-law emission which dominates the spectrum in those states, we compared the {\it RXTE} PCA spectra obtained for GRO~J1655-40 (high-inclination) and GX~339-4 (low-inclination) for two different values of hue: 85 degrees and 140 degrees.  The spectra can be well-described with a power-law and disc reflection component only, with no evidence for any disc blackbody contribution above 3~keV.  Furthermore, the reflection components (both iron line and Compton hump) appear slightly weaker in GRO~J1655-40 than in GX~339-4, so that the difference in hardness can only be explained if the power-law itself is systematically harder in GRO~J1655-40, at least over a range of hue from the hard state to the HIMS.

The inclination-dependence of the power-law spectrum could be explained if the corona has a flattened geometry, such that at higher inclinations, we look through a greater optical depth of Compton-upscattering material than at lower inclinations.  The fractional difference in hardness values between  40-150$^{\circ}$ in hue (where the spectrum is entirely power-law dominated) is on average 1.24. If we define the power-law spectral indices as $\alpha_{\rm hi}$ and $\alpha_{\rm lo}$ for the high and low inclination objects respectively, and obtaining the hardness ratio in terms of integrated fluxes for each band, we find that the corresponding difference is $\alpha_{\rm lo} - \alpha_{\rm hi} \approx 0.17$.  \citet{Pozdniakov79} show that if the electron temperature of the plasma $kT_{e}>50$~keV and the optical depth $\tau< 3$ then the power law index can be approximated by:
\begin{equation}
\alpha = -\frac{\ln(\tau) + \frac{2}{a+3}}{\ln(12a^2 + 25a)}
\end{equation}
Where $a = kT_{\rm e}/m_{\rm e}c^{2}$.  Given our approximate difference in the power-law indices between the high and low inclination objects and assuming a typical electron temperature within the corona of $kT_{\rm e}=100$~keV we can estimate the ratio of the two optical depths to be $(\tau_{hi}/\tau_{lo})=1.33$.  If we average the measured and limiting values of the inclination angles of our low and high-inclination sources, we find typical angles of 44$^{\circ}$ and 70$^{\circ}$ respectively.  For a plane-parallel geometry and uniform density (i.e. assuming that $\tau = \tau_{0}/\cos(\theta)$ where $\tau_{0}$ is the optical depth for a viewing angle of $\theta$=0$^\circ$) these angles imply a ratio of optical depths for high to low inclination of $\sim2.1$, which would be more than sufficient to produce the hardening we observe. We note that this interpretation implicitly assumes at most only a low level of misalignment between the hot inner flow and binary orbital plane. 

Of course, this model is very simplistic and does not account for the location of the seed photons and possibly linked radial temperature dependence (e.g. cooler on the outside) of the corona, but it is at least suggestive that a corona which is flattened in the plane of the disc could produce the inclination-dependent hardening which we see in the data.  Other mechanisms may also be at work, e.g. bulk motion Comptonisation in a hot inner flow \citep{Titarchuk98}, which could also produce a hardening of the spectrum seen closer to the direction of the flow velocity (i.e. in the disc plane).  By contrast, if the X-ray power-law is produced by Compton scattering in a jet, we would likely observe {\it softer} spectra for the high inclination objects \citep{Reig03}.  Therefore it seems more likely that at least for the HIMS and hard state with higher hue values, the power-law is not produced by scattering in a jet but in a flattened corona.  There is some hint however that the inclination-dependence disappears at lower values of hue, i.e. deeper into the hard state. This could indicate a transition to a less-flattened, even jet-like geometry or alternatively a more optically thin flow where fewer scatterings occur.

Finally, we comment on the evidence for inclination-dependence in the variability itself.  Inclination-dependence is seen in the amplitudes of the low-frequency QPOs, as noted by \citet{Schnittman06} and shown recently in a systematic way by \citet{Motta14}, who also note that the fractional rms of the broadband noise does not seem to depend on inclination.  We have strengthened this picture, by showing (Fig.~\ref{fig:pwcolinc}) that the power-colour-colour diagrams of low and high-inclination sources coincide once the QPOs are removed from the power spectrum.  Furthermore, we have shown (Fig.~\ref{fig:rmsincqporm}) that the rms vs. hue relation also does not depend on inclination.  Taken together, these results imply that while the QPO amplitude is stronger at higher inclinations, the broadband noise is independent of inclination both in amplitude {\it and} power-spectral shape.  This finding supports the idea that the physical origins of the QPOs and broadband noise are different, with the QPO likely to be a geometric effect, perhaps linked to precession of the inner hot flow or corona (e.g. \citealt{Ingram09}), while the broadband noise is likely to be associated with mass-accretion fluctuations intrinsic to the accretion flow and is probably well-correlated with the intrinsic structure of the central accreting and emitting regions.  Precession of the corona or inner hot flow will produce oscillations which increase with inclination angle if the corona is flattened in the plane of the disc.  Thus the inclination-dependence of both power-law slope and QPO amplitude may be deeply connected.

\section*{Acknowledgements}
We would like to thank the anonymous referee for their helpful comments and suggestions. This research has made use of data obtained from the High Energy Astrophysics Science Archive Research Center (HEASARC), provided by NASA’s Goddard Space Flight Center, and also made use of NASA’s Astrophysics Data System.


\bibliographystyle{mn2e}
\bibliography{powcolini.bib}

\bsp

\label{lastpage}

\end{document}